\documentstyle[12pt]{article}
\begin{document}
\def\beqra{\begin{eqnarray}} \def\eeqra{\end{eqnarray}}
\def\beqast{\begin{eqnarray*}} \def\eeqast{\end{eqnarray*}}
\def\beq{\begin{equation}}	\def\eeq{\end{equation}}
\def\be{\begin{enumerate}}   \def\ee{\end{enumerate}}
\def\fnote#1#2{\begingroup\def\thefootnote{#1}\footnote{#2}\addtocounter
{footnote}{-1}\endgroup}
\def\ut#1#2{\hfill{UTTG-{#1}-{#2}}}
\def\sppt{Research supported in part by the
Robert A. Welch Foundation and NSF Grant PHY 9511632}
\def\utgp{\it Theory Group,  Department of Physics \\ University of Texas,
 Austin, Texas 78712}
\def\lag{\langle}
\def\rag{\rangle}
\def\rta{\rightarrow}
\def\haf{\frac{1}{2}}
\def\om{\omega}

\def\ijmp#1#2#3{{\it Int. J. Mod. Phys.} {\bf A#1,} #2 (19#3) } 
\def\mpla#1#2#3{{\it Mod.~Phys.~Lett.} {\bf A#1,} #2 (19#3) }   
\def\npb#1#2#3{{\it Nucl. Phys.} {\bf B#1,} #2 (19#3) }  
\def\np#1#2#3{{\it Nucl. Phys.} {\bf #1,} #2 (19#3) } 

\def\plb#1#2#3{{\it Phys. Lett.} {\bf B#1,} #2 (19#3) } 
\def\prc#1#2#3{{\it Phys. Rev.} {\bf C#1,} #2 (19#3) } 
\def\prd#1#2#3{{\it Phys. Rev.} {\bf D#1,} #2 (19#3) } 
\def\pr#1#2#3{{\it Phys. Rev.} {\bf #1,} #2 (19#3) } 
\def\prep#1#2#3{{\it Phys. Rep.} {\bf C#1,} #2 (19#3) } 
\def\prl#1#2#3{{\it Phys. Rev. Lett.} {\bf #1,} #2 (19#3) } 
\def\rmp#1#2#3{{\it Rev. Mod. Phys.} {\bf #1,} #2 (19#3) }

\parindent=2.5em

\ut{06}{97}


\begin{center}

\Large{\bf Matrix Theory and U-duality in Seven Dimensions }\normalsize 

\vspace{24pt}
 Moshe Rozali\fnote{*}{\sppt} 

\vspace{12pt}
{\it \utgp}

\vspace{12pt}
\end{center}

\abstract{We demonstrate the emergence of the U-duality group in 
compactification of Matrix theory on a 4-torus. The discussion   
involves non-trivial effects in strongly coupled 4+1 dimensional gauge theory, and
highlights some interesting phenomena in the Matrix theory description of 
compactified M-theory.}

\baselineskip=21pt
\setcounter{page}{1}
\input epsf.tex
\def\theequation{\thesection.\arabic{equation}}
\section{Introduction}

\indent
   In the last few months there has emerged a candidate for the mysterious
 M-theory, formulated in the light cone gauge. The so called M(atrix)
 theory \cite{bfss}
 passed several consistency conditions.
 It has the right structure of 11-dimensional supergravity, a membrane and a
 longtitudal fivebrane \cite{bd}, with the right interactions \cite{int,bc}. In
\cite{bfss,sus,tori,bc,sus2}
the compactification on tori was considered, with emphasis on T-duality,
which is not manifest in the original formulation. Compactification on more
 complicated spaces was also considered \cite{more}. Additional issues are
addressed in \cite{bss}.

One of the checks that the Matrix theory has yet to pass is the emergence 
of the expected U-duality groups in various dimensions. This was considered
 in \cite{sus} where the $SL(2,Z) \times SL(3,Z)$ symmetry in 8 dimensions was shown to
hold, partially as a consequence of S-duality in $\cal N$=4 3+1 dimensional YM theory.
 Subsequently the $SL(2,Z)$ U-duality in 9 dimension was studied \cite{sus2}. 

In this letter we show the emergence of the expected U-duality group in 7
dimensions. This involves some non-trivial phenomena in strongly coupled SYM
theories in 4+1 dimensions, which have a natural interpretation in M-theory.
 We also comment on a puzzle regarding gravitational anomalies on the 
Matrix model's base space.

\section{ U-Duality in Seven Dimensions}
\indent \indent

We consider compactifying M theory on a 4-torus down to 7 dimensions.
 The theory has 32 supersymmetries and is unique: the only possible multiplet
 is the gravity multiplet. The scalars in that multiplet parametrize the
 manifiold $SL(5,R)/SO(5)$.
The conjectured U-duality group is $SL(5,Z)$. It is this structure that we want to discover
in the Matrix theory formulation of M-theory.

As was elaborated in \cite{tori,sus}, Matrix theory compactified on $T^d$ is described
by a d+1 dimensional field theory on the dual torus, with 16 supersymmetries.
The resulting theory for the case at hand is then a 4+1 dimensional 
theory with $\cal N$=2
SUSY. The vector multiplet of this theory has one vector, five scalars and two
Majorana spinors.
 The  matrix theory has  one vector multiplet in the adjoint of $U(N)$,
with the scalars representing the 5 non-compact transverse directions. For later use
 we prefer to use the $\cal N$=1 language, in which the $\cal N$=2 vector multiplet decomposes as
a vector multiplet (a vector, a spinor and a scalar) and an 
hypermultiplet (4 scalars and a spinor).

The action is
the usual SYM action:
\begin{equation}
L= \frac{1}{4g^2} tr F_{\mu \nu}F^{\mu \nu}+  ...         
\end{equation}
 
 The parameters of this theory are identified by a straightforward
application of the derivation in \cite{sus}.
This simple calculation is performed in the appendix. We take the space 4-torus
to be of sides $L_a$, and assume for simplicity it is rectangular. The gauge theory
then has base space of sides $S_a$ , and coupling constant $g^2$. Their values are given as:
\begin {eqnarray}
 S_a= \frac{2\pi (l_p)^3}{L_a R} \\
g^2=\frac{(2\pi)^2(l_p)^6}{R \, L_1 L_2 L_3 L_4}.
\end{eqnarray}
 Where $l_p$ denotes the 11-dimensional Planck length, and $R$ is the 
length of the 11th dimension, to be taken to infinity together with $N$.

 For general
parameters $L_i$ the effective coupling $Ng^2$ is not necessarily weak.
 Morover the SYM theory is not renormalizable, which would seem to lead to the 
problems of being not well-defined in the UV and being trivial in the IR. In
what follows we find a weakly coupled non-trivial fixed
point theory equivalent to this gauge theory.
 This makes the dynamics more transparent, and in particular the $SL(5,Z)$
becomes manifest.  

The gauge theory has a manifest $SL(4,Z)$ symmetry. It is the global symmetry mixing
 the parameters $S_a$, while keeping their product fixed. This doesn't change
 the gauge coupling (2.3). To extend this symmetry we look for another
parameter of the gauge theory to combine with the parameters $S_a$. The only
other parameter available is the gauge coupling itself. We note that 
$S_5 \equiv g^2$ has length dimension 1 in 4+1 dimensions, so it can
naturally combine with the parameters $S_a$. Furthermore, since the
coupling does not run, it can be treated as an additional parameter,
 independent
of the parameters $S_a$.

  We are therefore led to conjecture that the  Matrix theory 
description of this
 compactification is a 5+1 dimensional theory, with base space that is a 5-torus. The sides of the base space torus are the dual 4-torus,
 combined with $S_5$. 
The $SL(5,Z)$, then, is just
the geometric symmetry of the base space, mixing the parameters $S_i$ (i=1,...,5)
while keeping their product fixed.

 The excitations corresponding to momentum states in the additional dimension
are introduced in \cite{sei}. The gauge theory has instantons, which are point particles in 4+1
 dimensions \footnote 
{By an instanton, in any number of dimensions,
we mean a solution of the Euclidean YM equations with 4 transverse coordinates.}
, and have mass proportional to $\frac{1}{g^2}$ \footnote
{This is their classical mass. As BPS saturated states they do not receive
mass corrections.}.
 Analysis of the fermionic zero modes
reveals that these instantons are an $\cal N$=2 tensor (or equivalently a 
vector) multiplet in the adjoint of
$U(N)$ \footnote {Since the base space of the theory is a torus, there is no
non-compact collective coordinate corresponding to the instanton size. These are
zero size instantons.}.

 We propose the existence of a unique bound state
at threshold of $n$ of these instantons, for any $n$. We find support for this assumption
 in the next section, using known facts about type II string theory.
 As we take $S_5$  to
 be large , then, these states form a continuum, signalling 
a new dimension opening in the base space.

The origin of these states is also clear in M-theory: the instantons 
in the Matrix theory generally correspond to longtitudal fivebranes \cite{tori}, so the new
 states in the Matrix theory that carry momentum of the additional dimension are just
 fivebranes completely
wrapped around the 4-torus (and the lightcone direction). As an additional
evidence we show in the appendix that the instanton mass coincides
with the expected energy of a wrapped fivebrane. These states are
included in the gauge theory, in agreement
with the general approach to compactifications of the Matrix theory that
requires including all the 0-branes of the compactified theory \cite{bfss}
\footnote
{Note also that their mass is proportional to R, so they are easily interpreted
as particles with a definite $P_{11}$ \cite{bfss}.} .

 We see, then, that the wrapped fivebrane is included by the dynamics of the
 gauge theory, and does not have to be added by hand. Furthermore, T-duality
is incorporated naturally. As the spacetime torus shrinks to zero, we know
that the dynamics is more transparent in terms of the T-dual theory. 
In that theory the 4-branes are interchanged with the 0-branes. Therefore T-duality
predicts that for small values of $L_a$ the wrapped fivebrane should dominate 
the dynamics. This is indeed the case since this limit corresponds exactly to
strong coupling in the gauge theory, a limit in which the instantons 
become light. In the 5+1 dimensional description of the theory there is no
distinction between the original 0-branes and the wrapped 5-branes (instantons), thus
T-duality is manifest in that description.

 We conclude 
then that inclusion of the wrapped longtitudal fivebranes is accounted 
for by making the Matrix theory describing this compactification
 a 5+1 theory on a 5-torus. The U-duality group is then manifest classically
in Matrix theory as a global symmetry acting on the 
parameters $S_i$ \footnote{ See, however, the discussion below regarding
possible anomalies.}.
 
in the next section we  study the  strong coupling  behaviour of
the gauge theory by studying the 4-branes of type IIA theory, and demonstrate
 that the strong coupling limit is indeed a 5+1 dimensional theory. 
 
\section{Fourbranes in Type IIA Theory }
\indent \indent
 
 In this section we find support for the above conjectured behaviour of the
gauge theory by studying the 4-branes of type IIA theory.

The world volume theory of $N$ 4-branes is an $\cal N$=2 $U(N)$ gauge theory in 4+1
dimensions. It has 5 scalars in the adjoint, corresponding to the transverse
fluctuations of the brane. 
 The gauge coupling squared of this gauge theory is
  proportional
to  $\lambda$, the type IIA coupling.

In  M-theory
this 4-brane is the M-theory 5-brane wrapped around $X_{11}$, which is
of length $\lambda$. At strong coupling we see the 11th dimension
decompactify, and more relevant to our discussion, the gauge theory 
itself develops an extra 
dimension. In fact this configuration gives more information about the gauge theory:
at strong coupling the theory becomes the world-volume theory of $N$ coincident
 M-theory
5-branes, with  a chiral (2,0) SUSY in 5+1 dimensions.
 This is a  non-trivial theory and it
would be interesting to find in it more structure related to compactification
of M-theory to seven dimensions. The description of this theory as a limit
of a lower dimensional SYM theory might be useful since 
in the SYM theory there is no gauge invariant distinction between the
tensor multiplets describing the 5-branes (Cartan generators), and the tensionless strings that
connect the coincident 5-branes.  
 
 The theory in 5+1 dimensions has the required $SL(5,Z)$ as a classical symmetry.
 Since the theory is
chiral this symmetry can be anomalous. It was shown, in fact, that the M-theory's fivebrane
 world-volume theory has gravitational anomalies that can be canceled 
by anomaly inflow from the embedding space \cite{wit}. 
 Since this mechanism has no obvious interpretation in the present
context, it is not clear to us how the $SL(5,Z)$ symmetry survives quantum
corrections. This puzzle merits, in our opinion,  further investigation.

 The states on the 4-brane that carry momentum the extra dimension are,
 as disscused above, zero size instantons, which are 
identified in this context as the 0-branes of type IIA string theory. It is known that
there exist a unique bound state at threshold of $n$ such 0-branes and
 the 4-brane for every $n$ \cite{bound}. As the mass of these states is proportional
to $\frac{n}{g^2}$, they are naturally interpreted
 as Kaluza-Klein modes. As we take the limit of strong coupling the 4-brane becomes a 5-brane,
thus recovering the  M theory description.

 The existence of these bound states of $n$ 4-branes and $N$ zero branes has
 a somewhat simpler proof, since it is T-dual to the configuration of $N$ 4-branes and $n$
0-branes, a configuration for which the existence of a unique bound state
at threshold was established in \cite{bound}.
 
Therefore we conclude that our assumptions in the previous section about the
behaviour of the 4+1 dimensional theory at strong coupling have a  strong 
evidence in string theory \footnote {The emergence of a sixth dimensional
general covariance and chiral SUSY was disscussed  recently \cite{dis} for a 4+1 dimensional
theory with 16 supersymmetries.}. Due to possible anomalies, however, we
 can demonstrate the existence of the U-duality only as a classical
symmetry in the Matrix theory.

\section{Conclusions}
\indent \indent

As we compactify Matrix theory on more and more dimensions, more states have to
be considered in the theory, all of which decouple in the flat 
11-dimensional limit. First we have to add open strings
 connecting the 0-branes and wound around the compact dimensions. They
add dimensions to the base space of the gauge theory, according to the
 T-duality of type II string theory. As we compactify more than 2 dimensions
 there are also (closed) membranes wrapped around the compact dimensions.
 They are discovered in the the gauge theory as states (torons) that become
 light at strong coupling. As we compactify more than four dimensions
 there are also the longtitudal fivebranes, wrapped around the compact 
dimensions, that have to be considered. In this paper we have shown that
 these states
manifest themselves in the creation of an additional dimension in the Matrix 
theory's base space.

   As we compactify yet more dimensions we expect new phenomena to occur.
 In particular we expect to discover the effect of wrapped transverse fivebranes addeed
to the gauge theory.  The guide of the U-duality appearing as a symmetry of the
base space of the gauge theory, combined with some S-duality, might be useful
 in discovering the appropriate Matrix theory description of these compactifications (and the missing transverse fivebranes). 
 We hope to report on work on these topics in the near future.

\section*{Appendix}
\indent

 For the sake of completeness we repeat here the arguments of \cite{sus} 
in the case of compactification on a 4-torus. We take then the spacetime 
parameters to be $R$ and $L_a$ and the gauge theory's parameters to be $S_a$ and $g^2$.
 
  The energy of a singly wound string, that is a membrane wound around the direction $L_a$ and the 11th direction, is :
\begin{equation}
E=\frac {L_a R}{(l_p)^3}   .
\end{equation}
This is to be equated to the quantum of a momentum mode in the $S_a$ direction,
which is $\frac{2\pi}{S_a}$, therefore:

\begin{equation}
S_a= \frac{2 \pi (l_p)^3}{L_a R} .
\end {equation}

The homogenous modes of the gauge field have a term in the action:
\begin{equation}
 \frac{S_1 S_2 S_3 S_4}{2g^2} (\dot{A_a})^2
\end{equation}
which leads to energy quantum $\frac{(S_a)^2 g^2}{2 S_1 S_2 S_3 S_4}$ . This
is to be compared with the energy of a single 0-brane.
 The quantum of transverse momentum is $\frac{1}{L_a}$ and the quantum of 11th
dimension momentum is $\frac{1}{R}$.
Therefore the energy of the 0-brane is:
\begin{equation}
E=\frac {(P_a)^2}{2P_{11}}= \frac{R}{2(L_a)^2}
\end{equation}
This yields:
\begin{equation}
\frac{(S_a)^2 g^2}{S_1 S_2 S_3 S_4}= \frac{R}{(L_a)^2} .
\end{equation}
These two relations give:
\begin{equation}
g^2= \frac{(2\pi)^2 (l_p)^6}{ L_1 L_2 L_3 L_4 R} .
\end {equation}

 As a consistency check we calculate the energy of a 5-brane completely
wound around the 4-torus and the 11th dimension,it is:
\begin{equation}
E= \frac{L_1 L_2 L_3 L_4 R}{(l_p)^6}.
\end{equation}
 This is exactly the mass of a single instanton $\frac{4\pi^2}{g^2}$,
as claimed above.
 
\section {Acknowledgments}

The Author thanks D. Berenstein, R. Corrado, J. Distler,
 W. Fischler, S. Weinberg and especially M. Berkooz  and A. Hanany for useful conversations.


\begin{thebibliography}{99}
\bibitem{bfss} Tom Banks, Willy Fischler,
Stephen H. Shenker 
and Leonard Susskind,
hepth/9610043 .
\bibitem{sus2} Savdeep Sethi and Leonard Susskind, hepth/9702101.
\bibitem{bc} David Berenstein and Richard Corrado, hepth/9702108.
\bibitem{bound} S.Sethi and M. Stern, hepth/9607145;
Michael R. Douglas, Daniel Kabat, Philippe Pouliot and
Stephen H. Shenker, hepth/9608024.
\bibitem{sei} Nathan Seiberg, \plb {388}{753}{96}, hepth/9608111.
\bibitem{sus} Leonard Susskind, hepth/9611164.
\bibitem{bd} Micha Berkooz and Michael R. Douglas, hepth/9610236.
\bibitem{bss} Tom Banks, Nathan Seiberg and Stephen H. Shenker, hepth/9612157;
Gilad Lifschytz, hepth/9612223;
Miao Li, hepth/9612144;
Z. Guralnic and S. Ramgoolam, hepth/9702099. 
\bibitem{int} Ofer Aharony and Micha Berkooz, hepth/9611215;
Gilad Lifschytz and Samir D. Mathur, hepth/9612087.
\bibitem{dis} Jacques Distler and Amihay Hanany, hepth/9611104.
\bibitem{more} Michael R. Douglas, hepth/9612126;
Nakwoo Kim and Soo-Jong Rey, hepth/9701139;
Ulf H. Danielsson and Gabriele Ferretti, hepth/9610082;
Lubos Motl, hepth/9612198 ;
Shamit Kachru and Eva Silverstein, hepth/9612162.
\bibitem{tori} Washington Taylor IV, hepth/9611407;
Ori J. Ganor , Sanjaye Ramgoolam and Washington Taylor IV,
 hepth/9611202.
\bibitem{wit} Edward Witten, hepth/9610234.


\end{thebibliography}
\end{document}